# NUMERICAL SIMULATION OF TIDAL SYNCHRONIZATION OF THE LARGE-SCALE CIRCULATION IN RAYLEIGH-BÉNARD CONVECTION WITH ASPECT RATIO 1

[1]S. Röhrborn,* [1]P. Jüstel, [1]V. Galindo, [1]F. Stefani, [2]R. Stepanov

[1]*Helmholtz-Zentrum Dresden-Rossendorf, Institute of Fluid Dynamics, Department of Magnetohydrodynamics, Bautzner Landstraße 400, 01328 Dresden, Germany*
[2]*Institute of Continuous Media Mechanics, Russian Academy of Science, Acad. Korolyov St. 1, 614013 Perm, Russia*

*Corresponding author: s.roehrborn@hzdr.de

**Abstract**: A possible explanation for the apparent phase stability of the 11.07-year Schwabe cycle of the solar dynamo was the subject of a series of recent papers [1, 2, 3]. The synchronization of the helicity of an instability with azimuthal wavenumber m=1 by a tidal m=2 perturbation played a key role here. To analyze this type of interaction in a paradigmatic set-up, we study a thermally driven Rayleigh-Bénard Convection (RBC) of a liquid metal under the influence of a tide-like electromagnetic forcing. As shown previously, the time-modulation of this forcing emerges as a peak frequency in the m=2 mode of the radial flow velocity component. In this paper we present new numerical results on the interplay between the Large Scale Circulation (LSC) of a RBC flow and the time modulated electromagnetic forcing.

## 1. Introduction

There are quite a number of well-studied methods to stir liquid metals. The most prominent ones use traveling magnetic fields (TMF) to produce meridional flows [4] or rotating magnetic fields (RMF) to generate rotating flows. In combination, they can create "tornados" in liquid metals [5]. In contrast to the flow generation by TMF or RMF, we study here the flow that is caused, or influenced, by a tide-like forcing with its typical m=2 azimuthal dependence. Our main motivation to study such a forcing is related to the resonant excitation of intrinsic helicity oscillations of the kink-type (m=1) Tayler instability (TI) by tide-like (m=2) forces. Such a mechanism was recently invoked to explain the apparent phase stability of the solar Schwabe cycle in terms of synchronization with the 11.07-year spring-tide periodicity of the tidally dominant planets Venus, Earth and Jupiter [1,2,3].

In view of serious challenges to study the TI in a liquid metal experiment (which requires currents of a few thousand amps [6]), we consider here the Large Scale Circulation (LSC) of thermally driven Rayleigh-Bénard Convection (RBC) as a closely related flow with the very same azimuthal m=1 structure. This LSC is known to undergo sloshing and torsional oscillations [7], which are connected with helicity oscillations very similar to those of the very TI [8].

## 2. Numerical model

To investigate the influence of a tide-like forcing on the LSC of RBC, we consider an aspect ratio $\Gamma=1$ cylindrical container filled with the eutectic liquid metal GaInSn. The container with radius R=0.09 m is equipped with two thermally controlled copper plates (with thickness $h_{cu}$=0.025 m) at the bottom and top and an adiabatic cylinder wall. The tide-like force is imposed to the fluid by an electromagnetic field that is generated by two coils situated on opposite sides of the container (Figure 1).

The flow in the cell is computed by solving the incompressible Navier-Stokes equation and the continuity equation

$$\rho \frac{\partial \mathbf{u}}{\partial t} + \rho(\mathbf{u}\cdot\nabla)\mathbf{u} = -\nabla p + \mu\nabla^2\mathbf{u} + \mathbf{F}, \tag{1}$$

$$\nabla \mathbf{u} = 0, \tag{2}$$

$$\text{with } \mathbf{F} = M(t) * \mathbf{F}_{EM} + \mathbf{F}_{buoyancy} \tag{3}$$

with the open source library of OpenFOAM 6. The thermal convection is modelled by the Boussinesq approximation which provides the buoyant force term to the Navier-Stokes equation. In view of a factor 16 between the thermal diffusivities of copper and GaInSn we employ constant temperature conditions at the two interfaces at top and bottom.

The resulting flow velocities are in the order of a few mm/s, leading to small magnetic Reynolds numbers ($10^{-3}...10^{-2}$). In comparison with the characteristic speed of the applied AC magnetic field ($2\cdot\pi\cdot f\cdot D = 2\cdot\pi\cdot 25\cdot 0.18$ m/s = 28.3 m/s), the influence of the flow velocity onto the electromagnetic field is also negligible. These two facts considered jointly allows us to treat the driving force as flow-independent and hence to decouple the electromagnetic calculations in Opera 1.7 from the flow calculation in OpenFOAM. We therefore pre-computed the Lorentz force

$$\mathbf{F}_{EM} = \mathbf{j} \times \mathbf{B} \tag{4}$$

by solving the Maxwell equations and added it as a vector field to the Navier-Stokes equation. The resulting constant Lorentz force field amplitude-modulated by the factor

$$M(t) = \sin^2(t\cdot f_{LSC}\cdot\pi), \tag{5}$$

wherein the frequency $f_{LSC}$ is usually chosen to meet the resonance point for the interaction of the sloshing/torsional motion of the LSC and the tidal forcing (some other frequencies will also be tested). To achieve a positive sinusoidal modulation between $M(t) = 0$ and 1, the sinus is squared and the frequency is halved.

The generated mesh consists of hexahedral cells, with contracted cells at the walls where the no-slip condition u=0 is implemented. For all numerical simulations of the combination of RBC and tidal forcing, we used the results from a pure RBC after 3000 seconds as starting point.

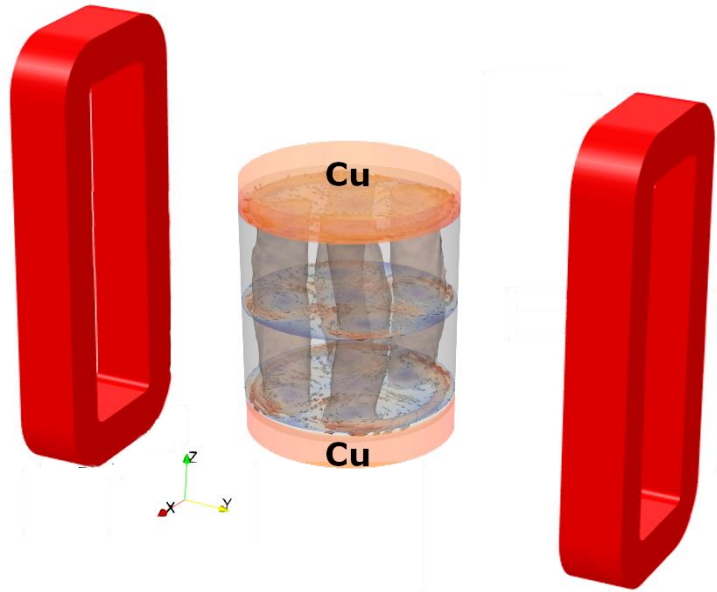

**Figure 1: Geometrical model with the example of a time averaged flow field produced by constant m=2 forcing without temperature difference**

## 3. Results and Discussion

The copper plates at top and bottom homogenize the force field in vertical direction which fosters a m=2 shaped force field without significant axial dependence (without the copper plates the electromagnetic force vectors would have a more complex structure, pointing both from the side wall as well as from top and bottom towards the container's center). Without any imposed temperature difference, a time-constant Lorentz force applied to a u=0 state generates four counter rotating vortices standing upright in the cylinder (see Figure 1). For this case, our previous paper [9] had shown a very good agreement between numerical and experimental results.

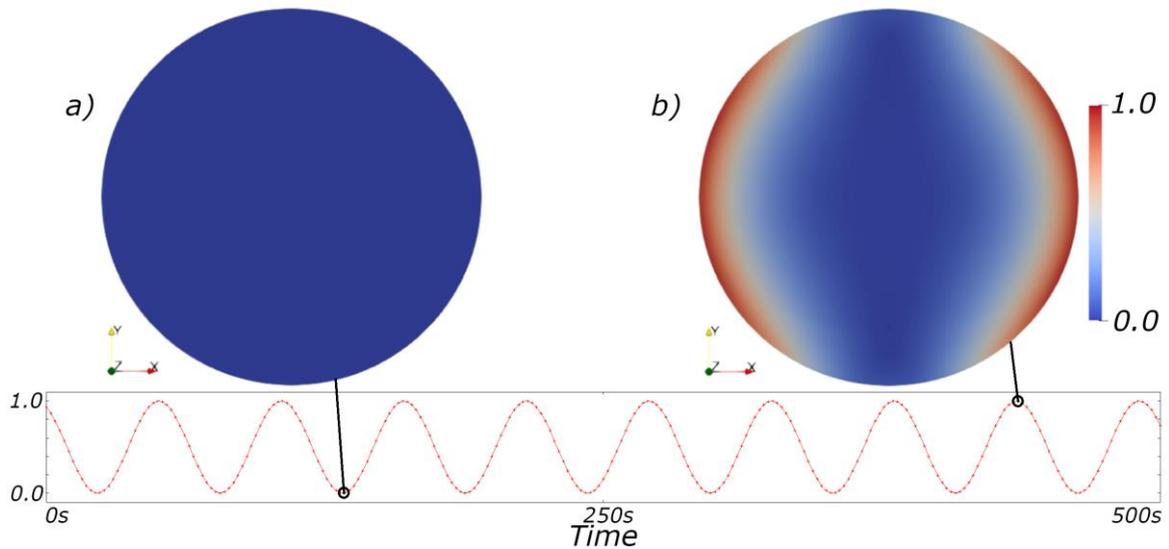

**Figure 2: Sinusoidal time modulation of the force field. Force distribution at minimum (a), at maximum (b) of the amplitude modulation according to Equation (5).**

Analyzing the flow thus generated reveals a clearly dominant m=2 mode. When implementing a slow modulation to this forcing with the modulation factor $M(t)$ according to Eq. (5) (Figure 2), the modulation frequency $f_{LSC}$ becomes clearly dominant in the spectrum of the m=2 mode of the radial velocity component, measured on a circle with radius 0.5*R in the central plane (Figure 3). This result serves as a starting point for the synchronization studies to be described in the following.

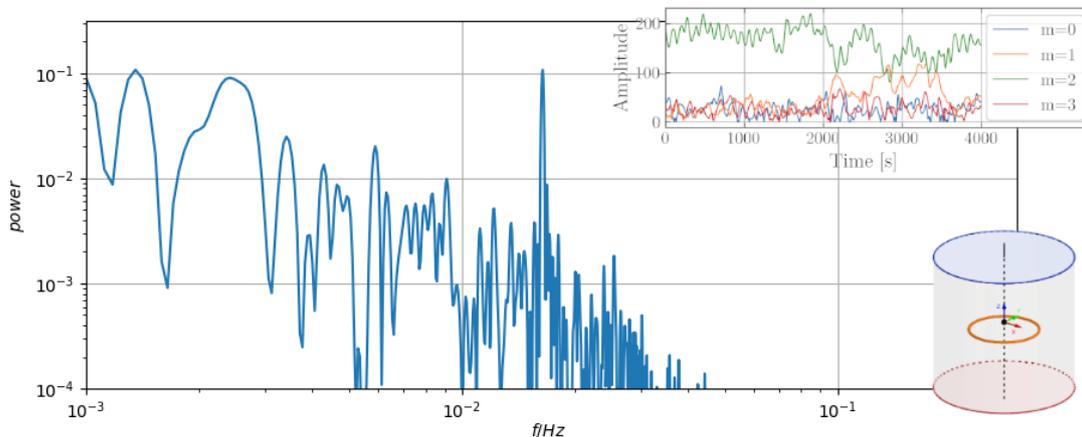

**Figure 3: Frequency spectrum of the m=2 mode of the radial velocity measured at sensor points on the orange ring. Note the clear peak at the modulation frequency of 0.01818 Hz.**

Before investigating the possible influence of a tide-like forcing on the sloshing/torsional motion of the LSC, we extracted the dominant frequency of the LSC for a pure RBC with a Rayleigh number of $1.03 \times 10^7$. That value corresponds to a temperature difference of ΔT=5K between the heated bottom plate and the cooled top plate. Both in the DNS simulation with a resolution of 547,992 cells and in the experiment (not shown here) we see a broad frequency band between 0.016 Hz to 0.021Hz, from which we chose the dominant oscillation frequency of $f_{LSC} = 0.01818 Hz$. Some comparative simulations with a higher resolution of up to $2 \times 10^6$ elements showed a similar behaviour in the macroscopic flow dynamics of the LSC. In view of the high numerical costs typically connected with the simulation of liquid metal convection [10], we decided to use the $0.55 \times 10^6$ elements mesh for all following numerical simulations.

When using the obtained frequency value for the modulation (according to Eq. (5)) of the tide-like force generated with a current of 30A and a frequency of 25 Hz, the slow meandering of the LSC can be stopped and locked into a well-defined band of angles (see Figure 4). Turning the force field by an angle of 60° locks the LSC to the new position turned by 60° (see Figure 5).

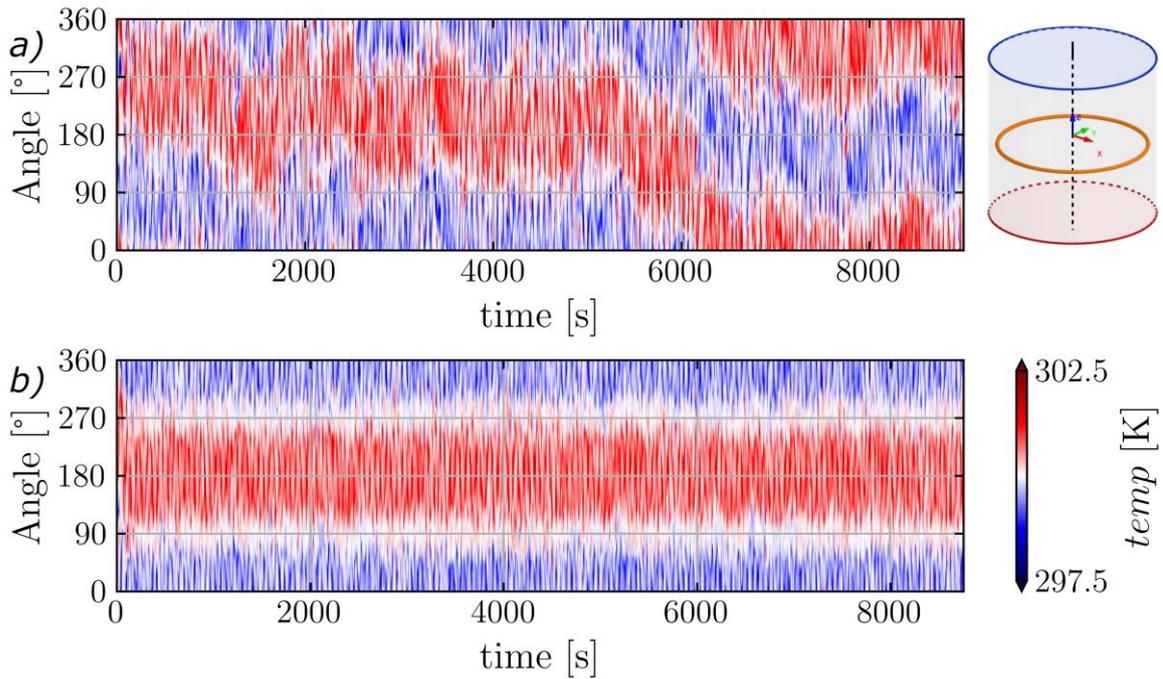

**Figure 4: Time series of the temperature distribution over angle taken from the orange sensor ring for a) a pure RBC and b) a RBC combined with a m=2 forcing (30A, 25Hz) and a modulation frequency of f$_{LSC}$=0.01818Hz. At t=0, the RBC is already fully developed (the prior transient over 3000 s is not shown here).**

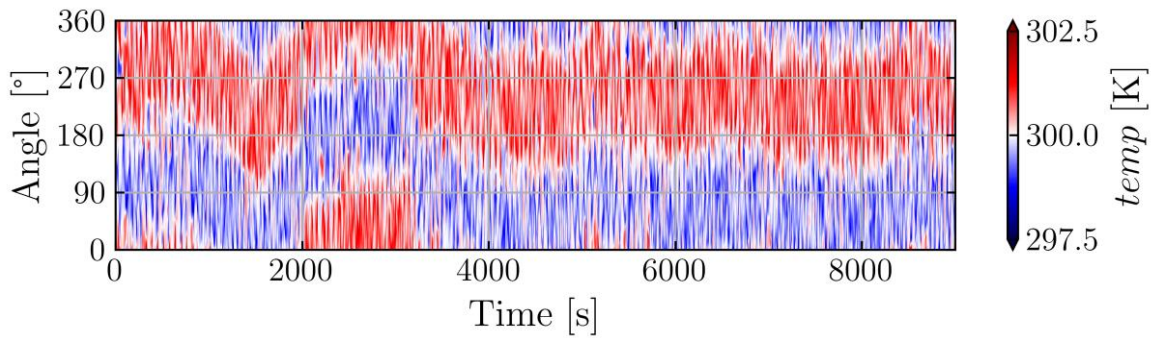

**Figure 5: Time series of the temperature distribution over angle for a RBC combined with a 60° turned m=2 forcing (30A, 25Hz) and a modulation frequency of f$_{LSC}$=0.01818Hz. At t=0, the RBC is already fully developed (the prior transient over 3000 s is not shown here).**

Analysing the mean helicity of the flow in the cylinder half -0.09 m < x < 0.0 m (Figure 6) shows that the broad frequency band (0.016Hz < f < 0.021Hz) of the oscillation in pure RBC (a) is drastically narrowed to 0.018 Hz under the influence of a m=2 force field, generated with a relatively high current of 27.6 A. The second peak at 0.024 Hz, which seems to correspond to a secondary 3:4 resonance, has still to be understood.

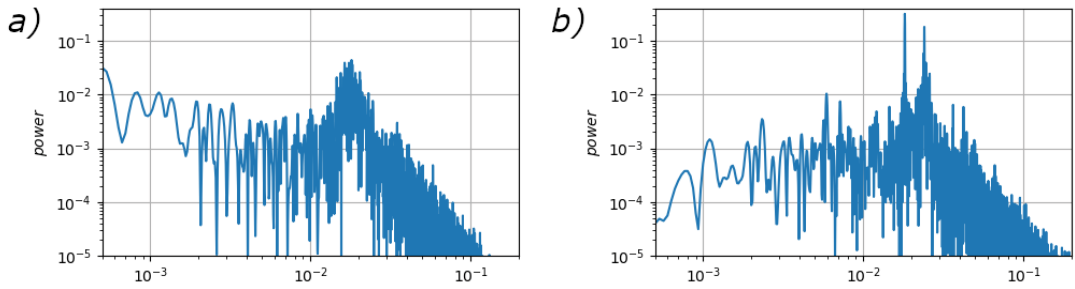

**Figure 6: Frequency spectrum of the mean helicity within -0.09 m ≤ x ≤ 0.0 m for a) a pure RBC and b) a RBC exposed to an additional m=2 forcing with 27.6A and 25 Hz and a modulation frequency of $f_{LSC}$=0.01818Hz.**

Figure 7 shows more details of the flow development when increasing the current. For a comparably low value of 5.9 A (Figure 7a), the LSC still has a rather irregular behaviour. For 12.5 A (Figure 7b), for which the typical flow speed due to the m=2 forcing acquires amplitudes of one third of the LSC, the flow becomes significantly more regular. For 27.6 A (Figure 7c), we observe a clear clocking of the m=1 LSC with the modulation of the tide-like m=2 forcing. The apparent clocking with the doubled frequency, which becomes evident in Figure 7d, has still to be analysed in more detail.

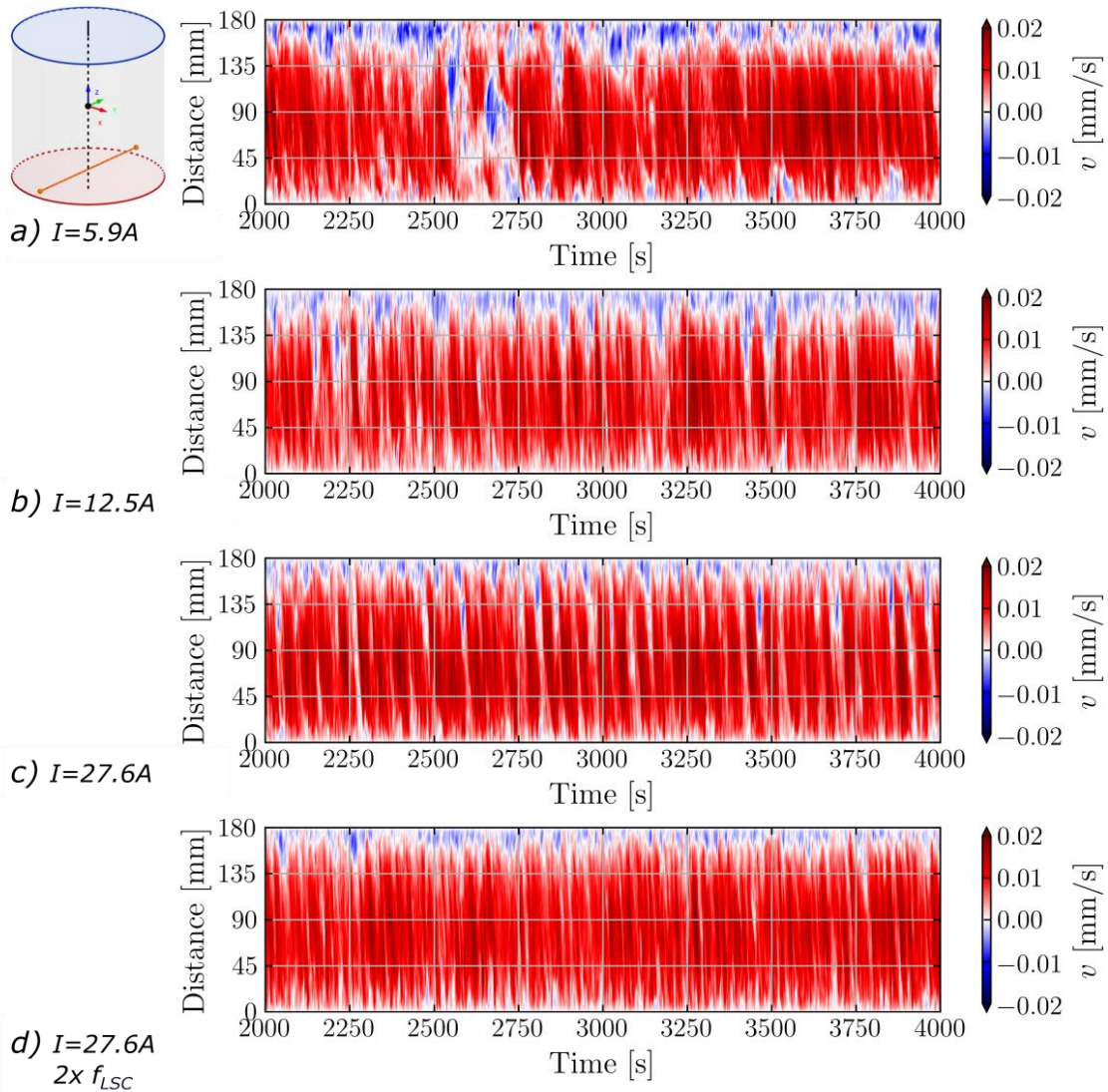

**Figure 7: Comparison of flow speeds taken from a sensor path (orange line) over time for different electromagnetic forces produced by different currents (f=25 Hz) and a modulation frequency $f_{LSC}$=0.01818 Hz (a,b,c) and 0.03636 Hz (d).**

## 4. Conclusion

We have studied the influence of a tide-like m=2 electromagnetic forcing on the flow structure of a RBC in an aspect ratio Γ=1 cylinder filled with the liquid metal GaInSn. Without any applied temperature gradient, the low-frequency sinusoidal modulation by the factor M(t) is clearly visible in the spectrum of the m = 2 flow mode.

The impact of the time-modulated force field onto the LSC oscillation depends on the strength of the imposed force. Force fields generated by an AC current of around 30A and a frequency of 25Hz have a significant influence on the flow structure. In particular, they lock the LSC, which under pure RBC condition undergoes a meandering motion, into a narrow band of angles. The sloshing/torsional motion of the LSC, and the helicity oscillation connected with it, becomes synchronized with the tide-like force.

In summary, we have shown that a m=2 tide-like forcing has a significant influence on the m=1 flow structure of an RBC. More numerical parameter studies and a confirmation of the results by experimental studies will be needed to establish the underlying synchronization mechanism. The critical current for the onset of synchronization remains also to be identified.

## 5. Acknowledgement

This work was supported in frame of the Helmholtz-Russian Science Foundation Joint Research Group "Magnetohydrodynamic instabilities," Contract Nos. HRSF-0044 and 18-41-06201(R.S.) and by the European Research Council (ERC) under the European Union's Horizon 2020 Research and Innovation Programme (Grant No. 787544).